\documentclass{moriond}

\def\Journal#1#2#3#4{{#1} {\bf #2}, #3 (#4)}

\usepackage{wrapfig}
\usepackage[dvipsnames]{xcolor}
\usepackage{amsmath,amssymb}
\usepackage{physics2}
\usephysicsmodule{ab} % automatic bracing

\makeatletter
\newcommand\vb{\@ifstar\boldsymbol\mathbf}
\newcommand\va[1]{\@ifstar{\vec{#1}}{\vec{\mathrm{#1}}}}
\newcommand\vu[1]{%
\@ifstar{\hat{\boldsymbol{#1}}}{\hat{\mathbf{#1}}}}
\makeatother

\newcommand{\transpose}[1]{\ensuremath{#1^{\mathsf T}}}

\newcommand{\Photo}{}

\begin{document}
\vspace*{4cm}
\title{Map-making Strategies For Next Generation CMB Polarization Experiments}

\author{ Simon Biquard }

\address{Université Paris Cité, CNRS, Astroparticule et Cosmologie\\ F-75013 Paris, France}

\maketitle\abstracts{
I briefly review some of the most common map-making strategies for experiments targeting the polarization of the Cosmic Microwave Background (CMB), in light of the anticipated volumes of data collected by next generation observatories such as Simons Observatory and CMB-S4.
Then, I focus on the pair-differencing approach, used for example in the POLARBEAR and BICEP collaborations.
Using simulations including a number of systematic effects, I evaluate the impact of correlated and unpolarized signals such as coming from the atmosphere, which is a major contaminant for ground-based experiments.
}

\section{Review of map-making flavors}

\emph{Map-making} is the reconstruction of the observed sky from the time-ordered data (TOD) collected by a telescope.
It compresses the volume of data by many orders of magnitude, while trying to preserve cosmological information.

The usual data model assumed for the map-making problem is:
\begin{equation}\label{eq:data_model}
\vb{d} = \vb{P} \vb{s} + \vb{n}
\end{equation}
where $\vb{d}$ is the (calibrated) TOD, $\vb{P}$ is the pointing matrix (which encodes the scanning and orientation of the telescope), $\vb{s}$ is the true sky map, and $\vb{n}$ is the noise vector.

Map-making is just a \emph{linear operation}, $\vb{m} = \vb{L} \vb{d}$, mapping the TOD to an estimator $\vb{m}$ of the true sky map.
In general, we can write \( \vb{L} = \pab{\transpose{\vb P} \vb{W_1} \vb{P}}^{-1} \transpose{\vb P} \vb{W_2} \), and the specific choices of $\vb{W_1}$ and $\vb{W_2}$ lead to different properties of the estimator.
A crude comparison of some possible combinations is shown in Table~\ref{tab:methods}.
We note that if $\vb{W_1} \neq \vb{W_2}$ the method may be biased, and this requires careful treatment at later stages of the analysis.
Also, the last row includes the standard destriping method which corresponds to the case where the filtered templates are piece-wise constant offsets.

\begin{table}[b]
\centering
\begin{tabular}{l l c c}
\hline
\textbf{Method} & \textbf{Choice} & \textbf{Pros} & \textbf{Cons}
\\ \hline
Binning
& $\vb{W_1} = \vb{W_2} = \vb{\Lambda}$
& unbiased, cheap
& complex noise
\\ \hline
GLS
& $\vb{W_1} = \vb{W_2} = \vb{C}^{-1}_{\vb n}$
& unbiased, min. variance
& expensive
\\ \hline
Filter-and-bin
& $\vb{W_1} = \vb{\Lambda} ; \vb{W_2} = \vb{F}$
& easy to compute
& biased
\\ \hline
Templates
& $\vb{W_1} = \vb{W_2} = \vb{F}$
& unbiased filtering
& expensive
\\ \hline
\end{tabular}
\caption{Comparison of different map-making approaches. Legend: $\mathbf{\Lambda}$ = diagonal noise weights; $\mathbf{C}^{-1}_{\mathbf n}$ = noise covariance matrix; $\mathbf{F}$ = filtering and weighting operator.}
\label{tab:methods}
\end{table}

These methods propose different trade-offs between precision, accuracy, and computational efficiency, and would need to be amended to mitigate some of the most pervasive systematic effects, which for the ground based experiments include \emph{atmospheric signal} and \emph{ground pickup}. In this work, we consider a hybrid approach, \emph{pair differencing}~\cite{bicep2-ii,polarbear}, which is well suited to accommodate such effects, could reach precision of the GLS approach, and can be computationally manageable.

\section{The pair differencing approach}

We assume that the atmospheric signal and the ground pickup are \emph{unpolarized}.
Then, for dual-polarization detectors $A$ and $B$, we can introduce a difference time stream:
\begin{equation}
\vb{d}_{A/B} = I \pm P
\quad\longrightarrow\quad
\vb{d}_- = \frac12 \pab{\vb{d}_A - \vb{d}_B} = P
\end{equation}
which is in principle free of atmospheric and total intensity sky signal. 
We use it to reconstruct the Stokes $Q$ and $U$ maps.

\begin{wrapfigure}{R}{0.5\textwidth}
\vskip -0.5truecm
\centering
\includegraphics[width=8truecm]{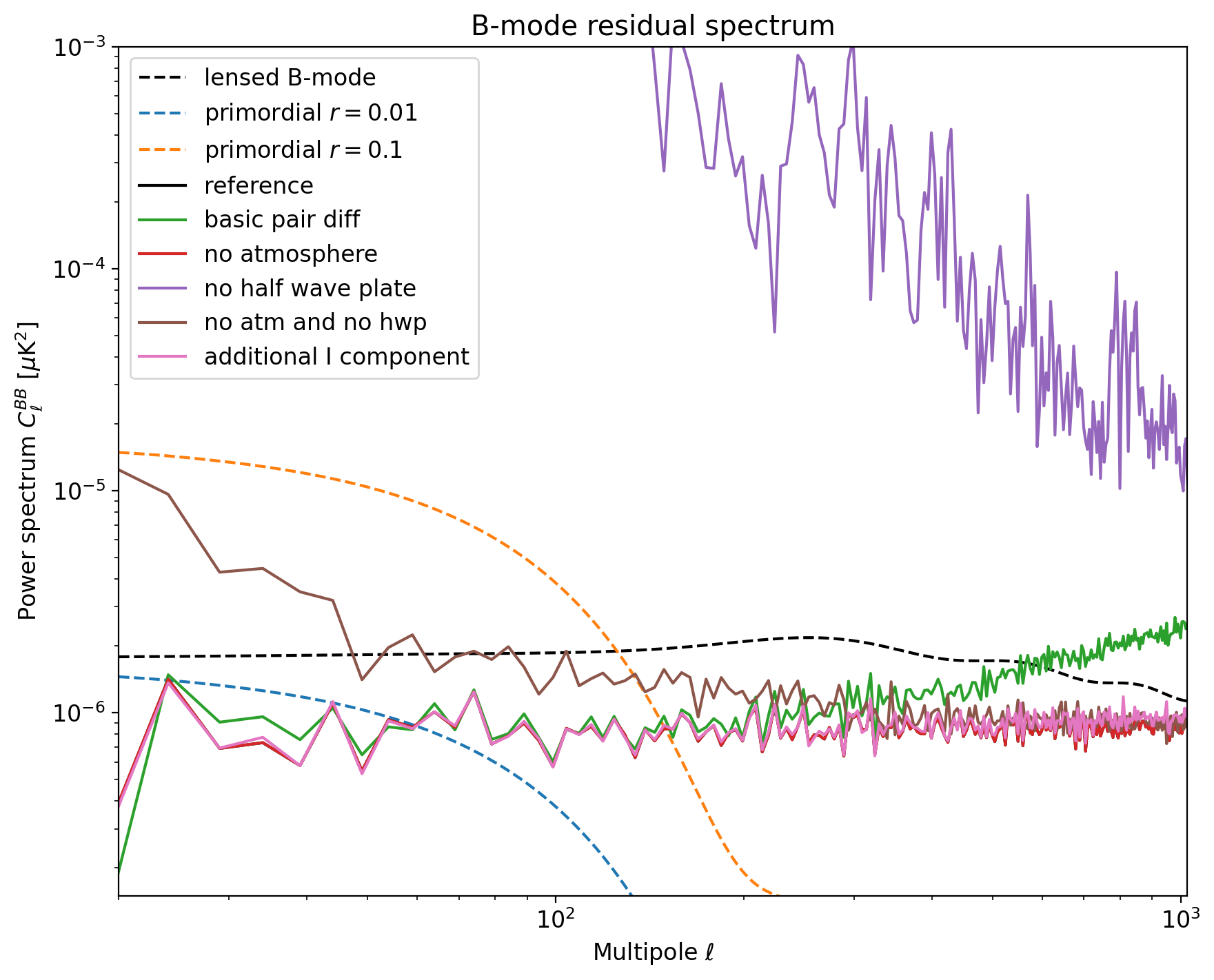}
\vskip -0.45truecm
\caption{
Dashed lines represent theory BB spectra (lensed in black, and primordial in blue and orange for a tensor-to-scalar ratio of 0.01 and 0.1 respectively).
Solid lines are the residual spectra for different configurations.
Except for the reference black curve, all runs have a one percent gain error.
\textbf{Black}: reference pair differencing (PD) case (hidden by red curve);
{\color{Green} \textbf{Green}}: PD run with atmosphere and HWP;
{\color{Red} \textbf{Red}}: PD run without atmosphere (aligned with black);
{\color{Purple} \textbf{Purple}}: PD run without HWP modulation;
{\color{Sepia} \textbf{Brown}}: PD run without atmosphere or HWP (increased residuals are due to instrumental noise which is not modulated);
{\color{VioletRed} \textbf{Pink}}: PD run, additionally fitting for the leaked intensity signal.
}
\vskip -1truecm
\label{fig:res}
\end{wrapfigure}

This approach removes unwanted, unpolarized signals, e.g., atmosphere, ground pickup, etc.; is numerically efficient, avoiding long-time noise correlations due to the atmosphere, and can produce maps with nearly optimal noise properties.
These advantages can only be exploited if the unwanted signals in both detectors are indeed \emph{identical}. This may not be the case due to, e.g., (i) miscalibration of the detector gains (ii) bandpass differences (iii) different beam shapes.
We here evaluate the impact of such effects.

\section{Preliminary results and outlook}

Our setup uses the Time-Ordered Astrophysics Scalable Tools (TOAST) package for simulating data and the MAPPRAISER library~\cite{mappraiser} for producing the sky maps.
We simulate Simons Observatory Small Aperture Telescope-like data with rotating half wave plate (HWP) and include the atmospheric noise.
In Fig.~\ref{fig:res}, we show a sample of our results displaying the \emph{residual spectra} of the difference between the reconstructed and input sky maps for a number of cases and compare them with targeted signals.

We conclude \emph{tentatively} that in the studied cases, the HWP helps mitigating the impact of the included systematics leading to an overall satisfactory performance of the method.
The on-going work focuses on (1) refining simulations by adding bandpass and beam differences; (2) inclusion of realistic gain calibration error distributions, and, (3) performing a likelihood analysis to estimate biases and their impact on the the tensor-to-scalar ratio.

\end{document}